\documentclass[a4paper]{jpconf}
\usepackage{graphicx}
\usepackage{subfigure}
\begin{document}
\title{Scintillator phase of the SNO+ experiment}

\author{V. Lozza$^1$ for the SNO+ collaboration}

\address{$^1$Technische Universit\"{a}t Dresden, Institut f\"{u}r Kern-und Teilchenphysik, Zellescher Weg 19, 01069 Dresden, Germany }

\ead{valentina.lozza@tu-dresden.de}

\begin{abstract}
The SNO+ experiment is the follow up of the SNO experiment, replacing the heavy water volume with about 780 tons of liquid scintillator (LAB) in order to shift the sensitive threshold to lower energy range. The 6000 m.w.e. natural rock shielding, and the use of ultra-clean materials makes the detector suitable for the detection of pep and CNO solar neutrinos, geo-neutrinos, reactor neutrinos and the possible observation of neutrinos from supernovae. Complementing this program, SNO+ will also search for $^{150}$Nd (5.6\% abundance) neutrinoless double beta decay, loading the liquid scintillator with 0.1\% of natural Neodymium. After a review of the general SNO+ setup, the physics of the solar neutrino phase will be presented.
\end{abstract}

\section{Introduction}

The unique location of the Sudbury Neutrino Observatory experiment 2 kilometres underground and the high radiopurity of the materials used in the construction of the detector, allowed to perform very accurate measurement of the spectral and flavour composition of the $^{8}$B solar neutrino flux \cite{SNO}. The SNO+ experiment is the follow up of SNO, in which the detector will be filled with 780 tons of liquid scintillator (LS). It will complete our understanding of the solar neutrino fluxes (complementary to SNO) lowering the energy threshold to \textit{pep} neutrinos (monoenergetic solar neutrinos flux of 1.442 MeV). Depending on the background level of U and Th chains and to the purity of the liquid scintillator itself, it will be also possible to detect \textit{pp} and \textit{CNO} solar neutrinos.\\
The Dirac or Majorana nature of the neutrinos fields will be also investigated using the SNO+ experiment by loading the liquid scintillator with 0.1\% of natural Nd, which contains $^{150}$Nd (5.6\% natural abundance), one of the neutrinoless double beta decay candidate isotopes. Having a high Q-value of 3.37 MeV \cite{150Nd}, the decay presents a low background due to the intrinsic radiopurity of the liquid scintillator.\\
During the pure liquid scintillator phase geo-neutrinos, reactor antineutrinos and supernovae neutrinos can also be measured.\\
In order to replace the heavy water by liquid scintillator, which has a lower density (0.86 g/cm$^{3}$) several detector developments were necessary. The most relevant changes are described in section \ref{Detector}. In section \ref{phase_of_operation} the phases of operation of the SNO+ experiment are summarized. The goals of the pure scintillator phase are then described in section \ref{scintillatore_phase}.

\section{Detector developments}\label{Detector}

The active volume of the SNO+ experiment will consist of 780 tons of Linear Alkyl Benzene (LAB) contained in a spherical acrylic vessel (AV) of 12 m diameter and 5 cm thickness, surrounded by about 9500 PMTs, for a coverage of 54\%. The PMTs and the AV are located in a cavity excavated in the rock filled with about 7000 tons of ultrapure water that acts as a first shielding layer, thermalizing neutrons coming from the rock. A radon seal and Urylon liner provide a second layer of shielding to avoid any contact of the detector with the mine air.
The 6000 m.w.e. rock shielding reduces the cosmic muon flux to about 3 muons per hour. \\
LAB was chosen as scintillator for the SNO+ experiment due to its chemical compatibility with acrylic, its high light yield (50 - 100 times higher than heavy water), the good optical transparency and low scattering. Moreover, high purity levels are easily available and due to the fast decay it is possible to discriminate between beta and alpha particles.\\
Heavy metal loading, such as Neodymium, is also being studied. The best solubility was found with organometallic compounds, however, the optical properties degrade with the increasing of the amount of metal dissolved. A good compromise was found for a loading of 0.1\% of natural Neodymium, corresponding to a total of about 44 kg of $^{150}$Nd in the entire detector volume.\\
Radiopurity requirements for the SNO+ experiment are very close to the purity levels reached by the Borexino experiment \cite{Borexino}. The level of Th and U should stay at 10$^{-17}$ g/g. In particular the level of $^{210}$Bi will determine if the measurement of \textit{CNO} solar neutrinos is possible. The shape of $^{210}$Bi is in fact very similar to the expected \textit{CNO} signal, making  the signal extraction very difficult if either the background levels present are too high or it is not possible to tag the background signal. 
The level of $^{14}$C is very important for the measurement of \textit{pp} solar neutrino flux. $^{14}$C, naturally present in liquid scintillator, emits a low energy beta that falls in the energy region of \textit{pp} neutrinos. If the level is too high it can completely hide the signal searched. Moreover, $^{14}$C can pile-up with other signals in the detector and it can fall in the neutrinoless beta decay energy region. Therefore, a ratio of $^{14}$C/$^{12}$C of 10$^{-18}$ is aimed for the SNO+ experiment.\\
Liquid scintillator will be purified with distillation, water extraction, nitrogen and steam stripping, and the use of metal scavenger adsorption columns. The system will be installed in the underground site in 2012.\\
From the detector development point of view, one of the main change is the installation of the new hold-down rope system. The SNO experiment used heavy water as active medium and therefore required the AV to be held up. On the other hand, SNO+ will use LS which, being less dense than water requires the AV to be held down in order to reduce the buoyancy force. An anchoring system was thus developed using Tensylon ropes, which have high purity level and thus very low background.\\
Before starting with the fill of the AV the inner surface must be polished and cleaned in order to remove Radon daughters, such as $^{210}$Pb, that can be plated on the surface during the exposure time to mine air after the removal of heavy water.\\
Replacement and reparation of the PMTs and the corresponding electronics must also be performed.\\
All these activities are schedule to be finished by the middle of 2012.\\
The PMTs will then be calibrated using both optical sources and beta-gamma sources.

\section{Phases of operation} \label{phase_of_operation}

Two phases of operation are planned for the SNO+ experiment.\\
One phase with pure liquid scintillator aim to study solar neutrinos, in particular \textit{pep} and \textit{CNO} neutrinos. Geo-, supernovae and reactor neutrinos are also part of the physics goals of this phase.\\
The second phase will be the Neodymium loaded phase. The main goal is the search for the neutrinoless double beta decay, however, the detector will remain sensitive to geo- and reactor neutrinos.

\section{Pure liquid scintillator phase}\label{scintillatore_phase}

After the precision measurement achived for the $^{8}$B and $^{7}$Be, to complete our understanding of solar neutrino fluxes the next target measurements are the \textit{pep}, \textit{pp} and \textit{CNO} neutrinos.

\subsection{\textit{pep} neutrinos}
\begin{figure}[h]
\subfigure{\includegraphics[width=17pc]{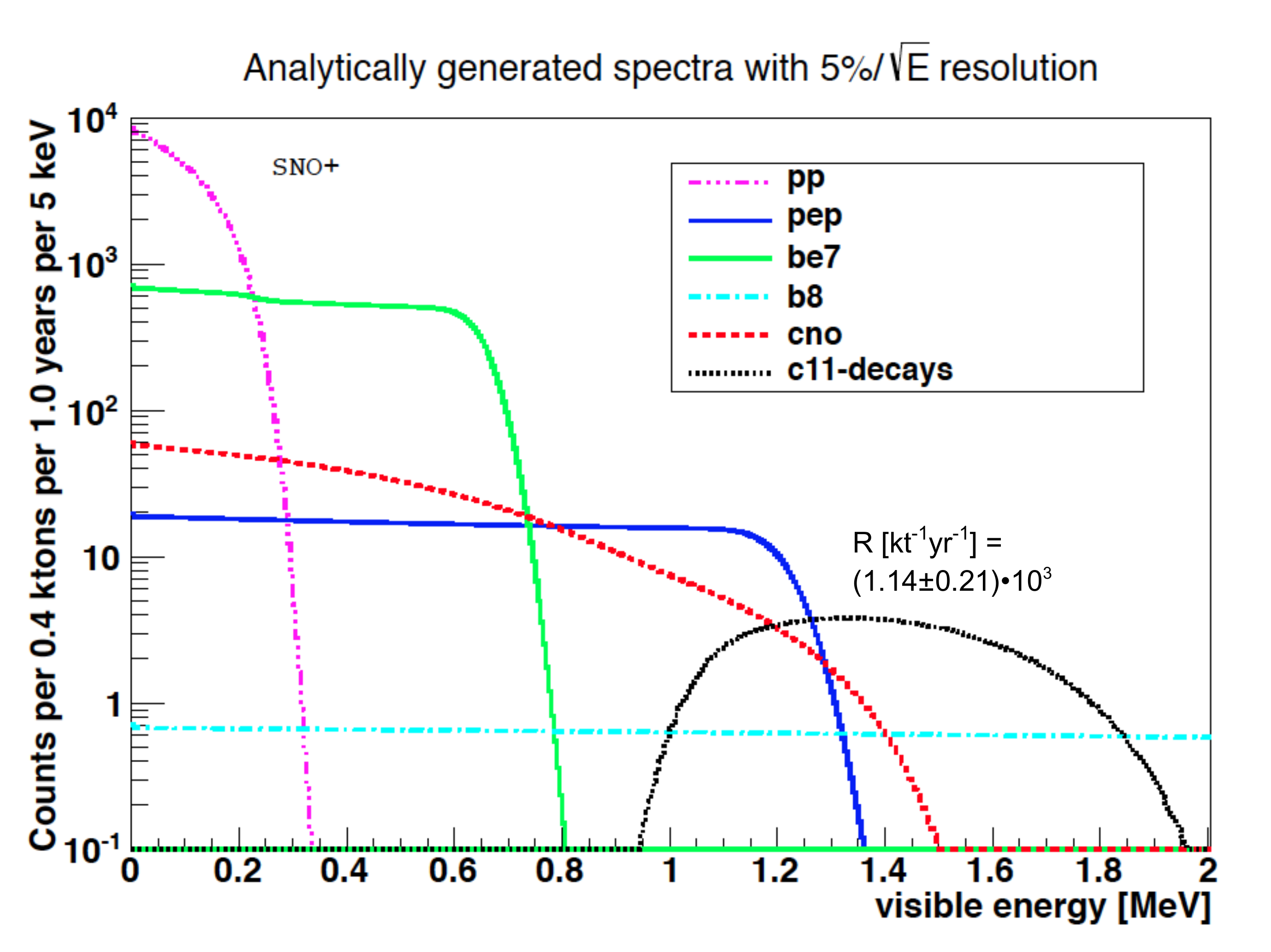}}\quad
\subfigure{\includegraphics[width=17pc]{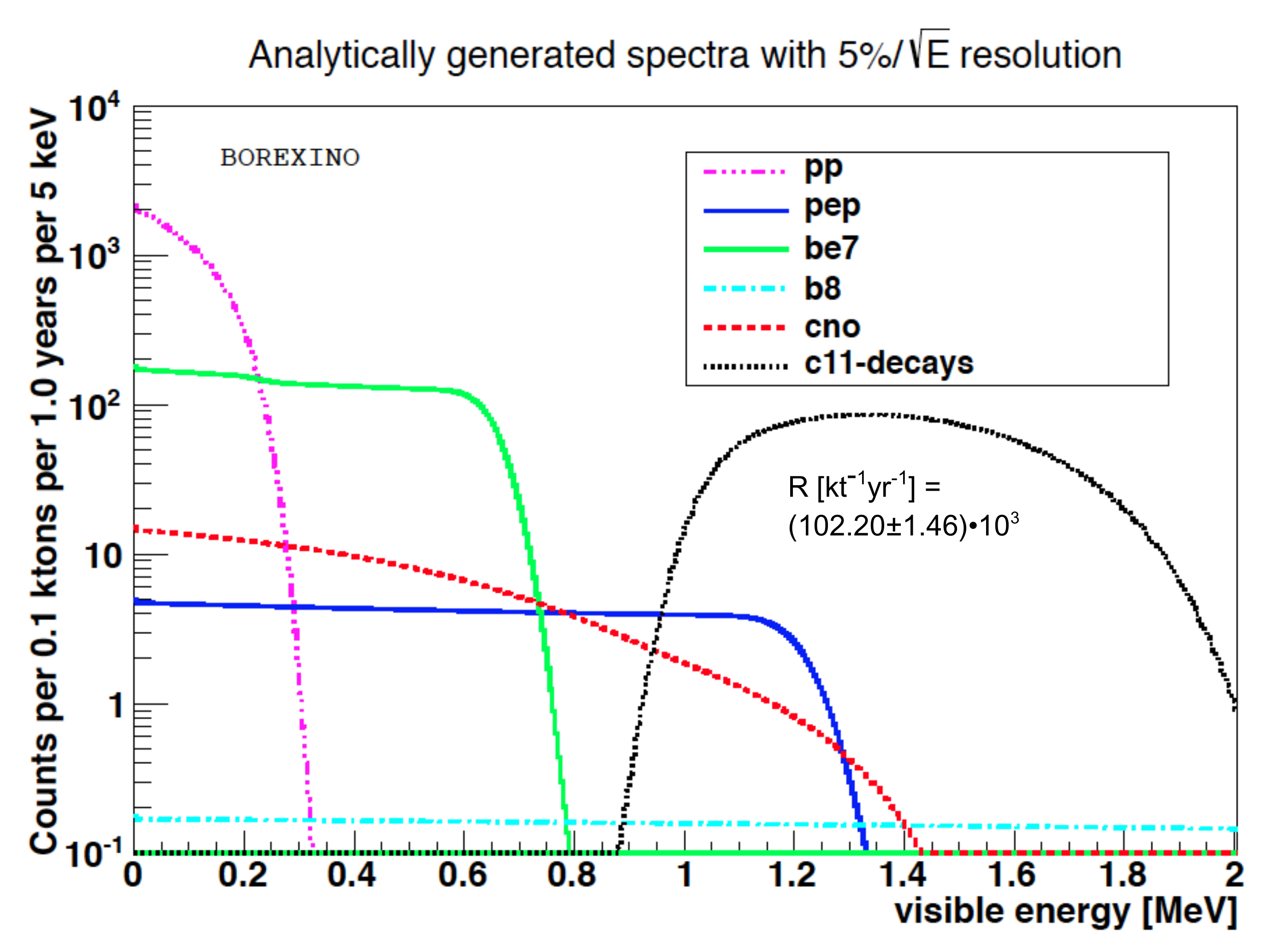}}
\caption{\label{11C} Simulated comparison between the expected $^{11}$C background in the SNO+ experiment (extrapolated from KamLAND data \cite{KamLAND}, left) and in the Borexino experiment \cite{Borexino_11C} (right) as a function of depth and fiducial volume for 1 year of measurement time. The resolution used is 5\%/$\sqrt{E}$. Other expected backgrounds are not shown.}
\end{figure}

The \textit{pep} reaction produces a monoenergetic neutrino (1.442 MeV) and has a very well predicted flux, with an uncertainty of 1.5\%, constrained by the Sun luminosity. An accurate measurement of the neutrino survival probability in this energy range can improve the precision on the active oscillation parameters and the sensitivity to alternative models of neutrino mixing.\\
In fig. 8 of \cite{Bellini} the survival probability of electron neutrinos as a function of energy is shown along with the most recent solar neutrino experimental results. The \textit{pep} line is placed in the transition region between the vacuum dominated (E$_{\nu} <$1 MeV) to the matter dominated oscillation (E$_{\nu} >$4 MeV), becoming an important probe for the MSW model as well as alternate model like the Non Standard Interaction (NSI). The present solar and atmospheric neutrino data, do not completely constrain any model and NSI oscillation solution can in fact successfully fit the data. The value of P$_{ee}$ for \textit{pep} neutrinos can then be a perfect test for different models \cite{Pena}.\\
The requirements for measuring the \textit{pep} solar neutrino flux are mainly three: depth, good light output of the liquid scintillator and radiopurity.\\
The depth plays a key role in the reduction of the cosmogenic muon flux. Muons interacting with the carbon atoms of liquid scintillator produce $^{11}$C that is the main background for the signal searched \cite{Galbiati}. Although coincidence events can be used to partially tag this background \cite{Galbiati}, a deeper location is more effective. In fig. \ref{11C}, a comparison between the $^{11}$C background in SNO+ and Borexino is shown.\\
For the measurement of the \textit{pep} neutrinos the main radiopurity requirements regards U and Th which should be at the level of 10$^{-17}$ g/g. Amongst the different daughters $^{210}$Bi is the most important since it partially falls in the region of interest.

\subsection{\textit{CNO} neutrinos}

Neutrinos from the CNO cycle can be used to test the solar metallicity. Improved models (2005) suggested about 30\% lower metallicity than the older but quite successful 1998 model. This broke the excellent agreement between the solar model calculations and helioseismology, opening the question if elements are homogeneously distributed in the Sun. CNO neutrinos are a direct probe since they can measure the metallicity of the core as is reported in \cite{Serenelli}.
The main source of background for the measurement of CNO neutrinos is $^{210}$Bi. Its shape is very similar to the searched signal one, making the discrimination very difficult. A high radiopurity level, as well as a Radon shielding and a tagging system are under study in order to reduce this background.

%%%%%%%%%%%%%%%%%%%%%%%%%%%%%%%%%%%%%

%% FIGURE %%

%%%%%%%%%%%%%%%%%%%%%%%%%%%%%%%%%%%%%

\subsection{Other physics goals}

The study of reactor anti-neutrinos and geo-neutrinos is also a goal during the pure scintillator phase. Reactor anti-neutrinos can be used to test neutrino oscillations, while geo-neutrinos are used to check the model of Earth heat production.\\
SNO+ is surrounded by 3 main reactors, which leads to a flux about a factor 5 smaller than KamLAND, but since the rate is dominated by only two baselines, the expected spectrum will have more distinct features allowing for a much easier identification of the oscillation peaks compensating the lower statistics.\\
Furthermore, the lower reactor neutrino flux will make it possible to measure the geo-neutrino flux. The flux is expected to be higher than the one observed by KamLAND and Borexino experiment due to the location of SNO+ in a thicker continental crust.\\
Supernovae neutrinos can also be measured, since the reactions on carbon nuclei will provide sensitivity to both neutral and charged currents. SNO+ is expected to be part of the Supernovae Early Warning System (SNEWS).\\
Additionally, the \textit{pp} neutrinos can be measured if the level of $^{14}$C with respect to $^{12}$C is low enough to allow a low electronic threshold and a signal to noise discrimination. The aim of $^{14}$C/$^{12}$C ratio of 10$^{-18}$ is foreseen for the SNO+ experiment.

\section{Conclusion}

SNO+ is the follow up of the SNO experiment, replacing the heavy water with liquid scintillator which, thanks to it high light yield, will allow to investigate the low energy region (E$<$3.5 MeV) making possible the detection of \textit{pep} and \textit{CNO} solar neutrinos.\\
Two phases of operation with different physics goals are planned: a pure liquid scintillator phase to search for the low energy solar neutrinos and a Nd loaded phase for the search of neutrinoless double beta decay.\\
Other exciting physics goals that will be covered are confirmation of oscillation with reactor anti-neutrinos, geo-neutrinos in a geologically-interesting location, supernovae neutrinos watch.\\
The main changes needed to adapt the existing SNO experiment to the use of liquid scintillator, will be completed by the middle of 2012. Liquid scintillator detector filling is expect to start early in 2013.

\end{document}